\documentclass[a4paper, 11pt]{article} 
\usepackage[right=1in,left=1in,top=1in,bottom=1in]{geometry}
\usepackage{graphicx} 
\usepackage{braket}
\usepackage{mathrsfs}
\usepackage[T1]{fontenc} 
\usepackage{amsmath}
\usepackage{pxfonts}
\usepackage[T1]{fontenc}
\usepackage{amssymb}
\usepackage{natbib}
\usepackage{epstopdf}
\usepackage{setspace}
\usepackage{enumitem}
\usepackage{sectsty}
\usepackage{wrapfig} 
\sectionfont{\large}
\bibpunct{(}{)}{,}{a}{}{,}
\usepackage{tcolorbox}
    \usepackage{pgfplots}

\usepackage{tikz}   
\usepackage{tikz-3dplot} 

\newcommand{\qed}{\nobreak \ifvmode \relax \else
      \ifdim\lastskip<1.5em \hskip-\lastskip
      \hskip1.5em plus0em minus0.5em \fi \nobreak
      \vrule height0.75em width0.5em depth0.25em\fi}

\usepackage{bbm}
\usepackage{MnSymbol}
\usepackage[all]{xy}

\usepackage{xcolor,colortbl,array,amssymb}

\usepackage{epigraph}
\makeatletter

\newlength{\bibitemsep}
\newlength{\bibparskip}
\let\oldthebibliography\thebibliography
\renewcommand\thebibliography[1]{%
  \oldthebibliography{#1}%
  \setlength{\parskip}{\bibitemsep}%
  \setlength{\itemsep}{\bibparskip}%
}

\title{\Large \textbf{The Wentaculus: \\ Density Matrix Realism Meets the Arrow of Time}} 

\author{Eddy Keming Chen\thanks{Department of Philosophy,  University of California, San Diego, 9500 Gilman Dr, La Jolla, CA 92093-0119. Website: www.eddykemingchen.net. Email: eddykemingchen@ucsd.edu  }}

\date{\textit{Dedicated to the memory of Detlef D\"urr}}

\begin{document}
\bibliographystyle{apa}

\maketitle 

\setlength{\bibitemsep}{.01\baselineskip plus .01\baselineskip minus .01\baselineskip}

\section{Introduction}

Two of the most difficult problems in the foundations of physics are (1) what gives rise to the arrow of time and (2) what the ontology of quantum mechanics is. They are difficult because the fundamental dynamical laws of physics do not privilege an arrow of time, and the quantum-mechanical wave function describes a high-dimensional reality that is radically different from our ordinary experiences. 

In this paper, I characterize and elaborate on the ``Wentaculus” theory, a new approach to time’s arrow in a quantum universe that offers a unified solution to both problems. Central to the Wentaculus are (i) Density Matrix Realism, the idea that the quantum state of the universe is objective but can be impure, and (ii) the Initial Projection Hypothesis, a new law of nature that selects a unique initial quantum state. On the Wentaculus, the quantum state of the universe is sufficiently simple to be a law, and the arrow of time can be traced back to an exact boundary condition. It removes the intrinsic vagueness of the Past Hypothesis, eliminates the Statistical Postulate, provides a higher degree of theoretical unity, and contains a natural realization of “strong determinism." I end by responding to four recent objections. In a companion paper, I elaborate on Density Matrix Realism.

\section{Conceptual Foundations}

I start by reviewing the conceptual foundations of Density Matrix Realism and the arrow of time that are necessary to formulate the Wentaculus in \S3. 

\subsection{Density Matrix Realism}

To understand Density Matrix Realism, let us start with a more familiar thesis---\textit{Wave Function Realism}. 
When we consider realist solutions to the quantum measurement problem, such as Bohmian mechanics, objective collapse theories, and Everettian quantum mechanics, it is natural to consider the quantum state of the universe as an objective feature of reality. Moreover, it is widely believed that it has to be a pure state, represented by a wave function. Let us formulate the thesis as follows:

\begin{description}
  \item[Wave Function Realism] The quantum state of the universe is objective; it has to be pure. 
\end{description}
This characterization of Wave Function Realism is broader than that of \cite{AlbertEQM} and \cite{ney2021world}. For them, Wave Function Realism carries a specific commitment to understanding the universal wave function as a physical field that lives on a vastly high dimensional ``configuration'' space, from which the ordinary 3-dimensional space is emergent. This is, however, not the only way to be a realist about the wave function. For example,   the multi-field interpretation, spacetime state realism, and  the nomological interpretation also count as versions of Wave Function Realism \citep{chen2019realism}. 

We often assume that the quantum state of the universe, if objective, must be pure.   In quantum mechanics,  mixed states are often used to represent reduced or statistical states, as expressions of entanglement with other systems or of our ignorance of the actual pure state. However, there is no compelling argument why the universe cannot be in a fundamental mixed state,  one that does not arise from entanglement or lack of knowledge. In fact, it is easy to formulate Bohmian mechanics, collapse theories, and Everettian quantum mechanics with a fundamental density matrix \citep{durr2005role, maroney2005density, allori2013predictions, wallace2012emergent}.
Let us  consider an alternative to Wave Function Realism, called \textit{Density Matrix Realism}:
\begin{description}
  \item[Density Matrix Realism] The quantum state of the universe is objective; it can be pure or impure. 
\end{description}
This thesis allows more choices of the quantum state of the universe; it can be pure or impure (a ``mixed state''). The freedom to use impure states is crucial to the Wentaculus, which I discuss in \S3. 

Let us clarify some key terms in both theses.  

(i) ``The'': it implies uniqueness. Both theses differ from the account considered by \cite{wallace2016probability} that the universe has two physical states at the same time: a fundamental pure state and a fundamental mixed state. 

(ii) ``Quantum state of the universe'': both theses are about the quantum state of the universe. It does not logically entail that  subsystem quantum states must be objective or that they must be pure (or impure). 

(iii) ``Objective'': it means that the universal quantum state corresponds to an objective feature of reality, is not merely epistemic (encoding lack of knowledge), or pragmatic (merely a useful instrument for calculations). The meaning of objectivity  is left open-ended, making room for different ways to be a realist \citep{chen2019realism}.


(iv) ``Must be pure'' vs. ``Can be pure or impure'': this is the only difference between the two theses. ``Must'' and ``can'' are  modal concepts. Wave Function Realism restricts universal quantum states to only pure ones, while Density Matrix Realism allows both pure and impure universal quantum states.  However, the latter is compatible with additional laws of physics (such as the Initial Projection Hypothesis) that make it physically impossible for the universe to be in a pure state. 



Are there modifications of Bohmian mechanics,  collapse theories, and Everettian quantum mechanics for which Density Matrix Realism is a natural framework? Yes, there are. They have been discussed, but not necessarily endorsed, by several authors in the foundational literature. For example, \cite{durr2005role} and \cite{allori2013predictions} have considered density-matrix realist versions of Bohmian mechanics, GRW theory, and Everettian quantum mechanics. See also \cite{maroney2005density} and \cite{wallace2012emergent}.
For example, in the Bohmian framework, we can evolve the fundamental universal density matrix $W$ by  the Von Neumann Equation:
\begin{equation}\label{VNM}
i \hbar \frac{\partial \hat{W}}{\partial t} = [\hat{H},  \hat{W}]
\end{equation}
particle configuration by a new guidance equation \citep{durr2005role}:
\begin{equation}\label{WGE}
\frac{dQ_i}{dt} = \frac{\hbar}{m_i} \text{Im} \frac{\nabla_{q_{i}}  W (q, q', t)}{ W (q, q', t)} (q=q'=Q)
\end{equation}
and distribute the initial particle configuration by: 
\begin{equation}\label{WBR}
P(Q(t_0) \in dq) =  W (q, q, t_0) dq
\end{equation}
This version of Bohmian mechanics satisfies equivariance just as the wave-function version does \citep{durr1992quantum, durr2005role}. In the Everettian framework, we can unitarily evolve the fundamental universal density matrix $W$ by the same von Neumann equation (\ref{VNM}), understand the emergent branching structure via decoherence, and apply  decision theory or self-locating probability to recover the Born rule \citep{ChenChua}.  We also have the option to add a separable fundamental ontology in spacetime by defining a mass-density field \citep{allori2013predictions}: 
\begin{equation}\label{mxt}
m(x,t) = \text{tr} (M(x) W(t)),
\end{equation}
with
$	M(x) = \sum_i m_i \delta (Q_i - x),$
and
$Q_i \psi (q_1, q_2, ... q_n)= q_i \psi (q_1, q_2, ... q_n).$
Finally,  in the GRW framework, we can interrupt the unitary evolution of the fundamental universal density matrix  $W$  with spontaneous collapses that occur at rate $N\lambda$ (where $N$ is the number of ``particles'' in the universe): 
\begin{equation}\label{collapse}
W_{T^+} = \frac{\Lambda_{I_{k}} (X)^{1/2} W_{T^-} \Lambda_{I_{k}} (X)^{1/2}}{\text{tr} (W_{T^-} \Lambda_{I_{k}} (X)) }
\end{equation}
with $X$ distributed by the probability density
$\rho(x) = \text{tr} (W_{T^-} \Lambda_{I_{k}} (x))$,
where the collapse rate operator is defined as
$\Lambda_{I_{k}} (x) = \frac{1}{(2\pi \sigma^2)^{3/2}} e^{-\frac{(Q_k -x)^2}{2\sigma^2}}$. As in the wave function version of GRW, we can define $W$-GRWm and $W$-GRWf with local beables mass-density field $m(x,t)$ and flashes $F$. 

 Each of the preceding theories posits  a fundamental universal density matrix with precise laws of nature. Given appropriate choices of the universal quantum state, each density-matrix theory is empirically equivalent to its wave-function counterpart,  so that they cannot be distinguished even in principle by experiment or observation \citep{chen2019quantum1}. I return to this issue in \S5.1.

\subsection{The Arrow of Time}

To appreciate the Initial Projection Hypothesis in the Wentaculus, we need to review a standard account about the arrow of time. Given the  reversibility of the fundamental dynamical laws, we must locate the origin of macroscopic irreversibility somewhere else. A proposal that has been influential in foundational literature posits a low-entropy boundary condition called the Past Hypothesis.\footnote{See \cite{albert2000time}, \cite{goldstein2001boltzmann},  \cite{callender2004measures, sep-time-thermo}, \cite{lebowitz2008time},  \cite{north2011time},  \cite{wallace2011logic}, \cite{loewer2016mentaculus},  \cite{goldstein2019gibbs}, and \cite{chen2020harvard}. For some criticisms, see \cite{winsberg2004can} and \cite{earman2006past}.} Roughly speaking,  if our universe started in the Past-Hypothesis region of the global state space, it will (with overwhelming probability) wander into states of higher entropy and eventually arrive at the thermodynamic equilibrium. 

For concreteness, I  focus on the Mentaculus theory of \cite{albert2000time, albert2015after} and \cite{LoewerCatSLaw, loewer2012two}, a particular version of the neo-Boltzmannian approach to the foundation of statistical mechanics. It contains two assumptions in addition to the fundamental dynamical laws. Following Albert and Loewer, let us call this package  \textbf{the Mentaculus}: 
\begin{enumerate}
\item \textbf{Fundamental Dynamical Laws (FDL):} the classical microstate of the universe is represented by a point in phase space
 (encoding the positions and momenta of all particles in the universe) that obeys $F=ma$, where $F$ encodes the classical interactions. 
 \item \textbf{The Past Hypothesis (PH)}: at a temporal boundary of the universe, the microstate of the universe lies inside $M_0$, a low-entropy macrostate that 
  corresponds to a small-volume set of points on phase space that are macroscopically similar. 
\item \textbf{The Statistical Postulate (SP)}: given the macrostate $M_0$, we postulate a uniform probability distribution with respect to the natural measure 
 over the microstates compatible with $M_0$.
\end{enumerate}

\noindent
Some comments:

(i) Loewer borrowed the name ``Mentaculus'' from the Coen Brothers movie \textit{A Serious Man} (2009).  It means the ``probability map of the universe.'' The Classical Mentaculus provides a probability assignment for every proposition formulable in the language of the classical phase space. If correct, it may account for all the temporally asymmetric phenomena and underly the objective probability in deterministic physics and the special sciences \citep{loewer2016mentaculus}.  

(ii) There is a certain degree  of vagueness in the partition of state space into macrostates, and hence in PH and SP.  A macroscopic description of the initial state does not correspond to any exact region in phase space. Any choice of an exact region risks a certain kind of objectionable arbitrariness \citep{chen2018NV}.

(iii)  The probability distribution in SP can be regarded as an objective notion of ``most'' with which we can ignore the anti-thermodynamic initial microstates in the PH region of the state space. This can be interpreted as a kind of deterministic objective probability \citep{loewer2001determinism} or typicality measure \citep{goldstein2012typicality}. In the quantum case, it is distinct from and in addition to the Born rule postulate.  

How should we implement the Mentaculus in quantum theory? On Wave Function Realism, the natural strategy is to  replace the classical state with a quantum pure state. This is a standard picture of Boltzmannian quantum statistical mechanics \citep{goldstein2010approach}. We posit the quantum state of the universe as represented by a unitarily evolving  wave function (obeying the Schr\"odinger equation) that started out in a low-entropy region in the Hilbert space, represented by a low-dimensional Past-Hypothesis subspace that corresponds to low quantum Boltzmann entropy. We further postulate a uniform probability distribution over all wave functions compatible with the PH subspace, with respect to the natural surface area measure on the unit sphere of the subspace. Call this the \textbf{Wave-Function Mentaculus}. 

What about on Density Matrix Realism? A similar strategy is to replace the universal wave function with a universal density matrix that can be pure or impure. We  postulate a unitarily evolving  density matrix (obeying the von Neumann equation) that started out in the PH subspace. We further postulate a uniform probability distribution over all density matrices compatible with the PH subspace, with respect to the natural measure on the space of all such density matrices \citep{chen2022uniform}. Call this the \textbf{Density-Matrix Mentaculus}. 


An interesting feature of Density Matrix Realism is that there is another and perhaps more compelling way to implement the key idea. It satisfactorily removes the inherent vagueness of PH and eliminates the SP. That is the \textbf{Wentaculus}. 

\section{The Wentaculus}


\subsection{The Initial Projection Hypothesis}

Can we do better than the preceding quantum versions of Mentaculus, in the sense of obtaining a unique initial quantum state? I suggest that we can. Recall that for any finite-dimensional Hilbert space $\mathscr{H}$ there is a natural density matrix in that Hilbert space---its normalized projection operator $\frac{I}{dim \mathscr{H}},$  where $I$ is the identity / projection operator on $ \mathscr{H}$ and $dim \mathscr{H}$ is the dimension of $\mathscr{H}$. Moreover, in general $\frac{I}{dim \mathscr{H}}$ is the simplest object one can associate with $\mathscr{H}$, containing no more information than is contained by $\mathscr{H}$ itself. Hence, if $\mathscr{H}$ is simple to characterize, then $\frac{I}{dim \mathscr{H}}$ is also simple to characterize. 

As a special case, consider the particular Past-Hypothesis subspace $\mathscr{H}_{PH}$ (which, among other things, has very low dimension and thus  very low quantum Boltzmann entropy $S_B (\mathscr{H}) = k_B dim \mathscr{H}$). There is a natural density matrix in  $\mathscr{H}_{PH}$,  namely $\frac{I_{PH}}{dim \mathscr{H}_{PH}}$, with $I_{PH}$  the identity / projection operator on $ \mathscr{H}_{PH}$ and $dim \mathscr{H}_{PH}$  the dimension of $\mathscr{H}_{PH}$. It is as simple to characterize as the Past-Hypothesis subspace itself. Therefore, if the PH is sufficiently simple to be considered a law, then the natural density matrix  $\frac{I_{PH}}{dim \mathscr{H}_{PH}}$ is too. 

I propose \citep{chen2018IPH} the following posit about the initial density matrix of the universe,  called the \textit{Initial Projection Hypothesis} (IPH):

\begin{equation}\label{PHID}
\hat{W}_{IPH} (t_0) = \frac{\mathbb{I}_{PH}}{dim \mathscr{H}_{PH}},
\end{equation}
 All the arguments that the PH should be nomological apply to IPH. I think the best understanding of this posit is a fundamental law of nature. After all, it is no more complicated and no less informative than usual versions of the PH  \citep{chen2020harvard}. 


The  posit can be generalized to other types of initial constraints. Here is a recipe:  starting from the full Hilbert space (energy shell) $\mathscr{H}$, we can use simple principles (if there are any) to determine an initial subspace $\mathscr{H}_0 \subset \mathscr{H}$, choose the natural quantum state in that subspace---the normalized projection $\hat{W}_{0} (t_0) = \frac{I_{0}}{dim \mathscr{H}_{0}}$. The natural choice will be simple and unique.

\subsection{Three Versions of the Wentaculus}

When we add the IPH to Density Matrix Realism, we arrive at the Wentaculus\footnote{The Wentaculus is so named because (1) it is inspired by the Mentaculus, and (2)  ``W'' is sometimes used to denote the fundamental density matrix. }:

\begin{enumerate}
\item \textbf{Fundamental Dynamical Laws (FDL):} the quantum state of the universe is represented by a density matrix  $\hat{W}(t)$ that obeys the von Neumann equation (\ref{VNM}).\footnote{For GRW-type theories, the density matrix obeys the stochastic modification of the von Neumann equation described in footnote \#22. }
\item \textbf{The Initial Projection Hypothesis (IPH)}: at a temporal boundary of the universe, the density matrix is the normalized projection onto $\mathscr{H}_{PH}$,  a low-dimensional subspace of the total Hilbert space. (That is, the initial quantum state of the universe is $\hat{W}_{IPH} (t_0)$ as described in equation (\ref{PHID}).)
\end{enumerate}
The Wentaculus implements the key idea of the Mentaculus for quantum theory. However, unlike the the Wave-Function Mentaculus, it requires Density Matrix Realism; moreover, unlike the Density-Matrix Mentaculus, its boundary condition law narrows down the choices of the initial density matrix to a unique one.
It contains one fewer fundamental postulate than each of the preceding versions of the Mentaculus. The Statistical Postulate becomes redundant, because there is exactly one nomologically possible initial density matrix. In  earlier versions of the Mentaculus, there are infinitely many nomologically possible initial states compatible with the PH. When we replace PH with IPH, we have only one initial state left. 

The Wentaculus is compatible with realist solutions to the quantum measurement problem.  For the Bohmian Wentaculus, we postulate that the state of the universe is described by particle configuration and the universal density matrix, and we add the IPH to the list of fundamental laws,  described by equations (\ref{VNM}), (\ref{WGE}), and (\ref{WBR}). Given the IPH, the initial quantum state of the universe is nomologically necessary. Hence,  there is only one nomologically possible velocity field for the particle configuration.  This differs from the Bohmian Mentaculus or standard versions of Bohmian mechanics, where there is nomological contingency about the initial quantum state. 

For the Everettian Wentaculus, we postulate that the state of the universe is described by the universal density matrix, and we add the IPH to the deterministic dynamical law,  described by equation (\ref{VNM}). Given the IPH, the initial quantum state of the universe is nomologically necessary, rendering the theory strongly deterministic \citep{chen2022strong}. Given the fundamental laws, there is only one possible history of the universal density matrix, and hence only one possible history of the Everettian multiverse. I return to this issue in \S4.3. 

For the GRW Wentaculus, we add the IPH to the stochastic dynamical law, described by equation (\ref{collapse}). Given the IPH, the initial quantum state of the universe is nomologically necessary, but because of the stochastic dynamics, the theory is not strongly deterministic. There are many nomologically possible histories of the universal quantum state, corresponding to different collapse histories permitted by the theory. 

Each Wentaculus theory is empirically equivalent to its Mentaculus counterpart, but they have physically inequivalent sets of models. For example, in the Bohmian case, the two theories will yield different particle trajectories. In the Everettian case, the two theories will yield different multiverses with different branches and different local descriptions.

\subsection{Realist Interpretations of the Density Matrix}

The Wentaculus is compatible with realist interpretations of quantum mechanics and of the quantum state. 


We have  four ontological interpretations of the universal density matrix. 
First, we can understand $W(q,q',t)$ as representing a physical field evolving in a $6N$-dimensional fundamental space represented by $\mathbb{R}^{6N}$. The field assigns properties to every point on that space.  
  Second, we can understand it as representing a low-dimensional multi-field. The fundamental space is a $3$-dimensional space represented by $\mathbb{R}^{3}$, and $W(q,q',t)$  assigns properties (represented by complex numbers or vectors) to every $2N$-tuple of points on that space. 
  Third, we can understand it as representing properties of spacetime regions. We can obtain, from the universal density matrix, reduced density matrices that correspond to physical properties of  regions in a $4$-dimensional manifold. Such properties are in general non-separable due to quantum entanglement. 
  Finally, we can understand it as representing a geometric object in Hilbert space. 
The Wentaculus is also compatible with the nomological interpretations of the quantum state. Moreover, as I explain in \S4.1, it solves the problem of complexity. While a generic universal quantum state of both quantum versions of  Mentaculus is enormously complicated, the initial density matrix postulated by the Wentaculus is sufficiently simple to be a law.  For more details on these realist interpretative options, see \cite{chen2019realism}. 



Of course, we need not be realists to accept the Wentaculus. There are non-realist quantum interpretations   according to which the quantum state represents our knowledge (or the lack thereof) or practical guidance for how we should act. QBists and quantum pragmatists can regard the Wentaculus as giving them the best epistemic or practical guidance for what to believe and what to act. In fact, they may be more comfortable with mixed states than some realists are. 


\section{Implications}

The Wentaculus has implications for several debates in foundations of physics. 

\subsection{The Nature of the Quantum State}

On the Wentaculus, we have the option to regard the quantum state as ontological. 
However, we also have an improved option for the nomological interpretation of the quantum state. As already mentioned, if PH is sufficiently simple to be a law, then the normalized projection onto the PH subspace is sufficiently simple. 

What does it mean to say the quantum state is nomological? As an analogy, consider the standard suggestion that the Hamiltonian function in classical mechanics is nomological. In the Hamilton's equations: 
\begin{equation}\label{HE}
\frac{d q_i}{d t} = \frac{\partial H}{\partial p_i} \text{  ,  } \frac{d p_i}{d t} = - \frac{\partial H}{\partial q_i},
\end{equation}
the $q_i$ and $p_i$ obviously represent something in the ontology. They have the ``marks of the ontic.'' They take on complicated values, and their values are not completely fixed by the theory and thus nomologically contingent.  
In contrast, the Hamiltonian function $H$ is very different: $H$ generates motion; $H$ is simple; and $H$ is fixed by the theory (and nomologically necessary). 
According to the standard interpretation,  H is not ontological but nomological. It does not represent things like particles or fields but a law that tells particles and fields how to move. 

Consider the guidance equation in Bohmian Wentaculus, with the right hand side expanded with a fixed initial density matrix $W_0$: 
\begin{equation}\label{N4}
\frac{dQ_i}{dt} = 
\frac{\hbar}{m_i} \text{Im} \frac{\nabla_{q_{i}}   \bra{q} e^{-i \hat{H} t/\hbar} \hat{W}_{IPH} ( t_0) e^{i \hat{H} t/\hbar} \ket{q'} }{ \bra{q} e^{-i \hat{H} t/\hbar} \hat{W}_{IPH} (t_0) e^{i \hat{H} t/\hbar} \ket{q'}} (q=q'=Q)
\end{equation}
$W_0$ has a similar character as the Hamiltonian function in Hamilton's equations:  $W_0$ generates motion; $W_0$ is simple; and  $W_0$ is fixed by the theory (nomologically necessary).
 $W_0$ has the marks of the nomic and \textit{can} be given a nomological interpretation.
This is to be contrasted with the Bohmian Mentaculus: its initial quantum state (either pure or mixed) is not guaranteed to be simple by the PH.  Implementing the nomological interpretation would require a different  argument,  perhaps by appealing to considerations about quantum gravity \citep{goldstein1996bohmian, goldstein2001quantum}. 
 
 The nomological interpretation can also apply to certain versions of the Everettian Wentaculus with a mass-density ontology. We can understand that the mass density is constrained by a law: 
 \begin{equation}\label{Nmxt}
m(x,t) = \text{tr} (M(x) e^{-i \hat{H} t/\hbar} \hat{W}_{IPH} ( t_0) e^{i \hat{H} t/\hbar})
\end{equation}
 Here $W_0$  does not generate motion in the sense of giving a velocity field, but it still generates the exact shape of the mass-density field in ordinary spacetime. Since $W_0$  is simple, we can take this equation as the fundamental law in the Everettian theory with a mass-density ontology, and $W_0$ is again nomological. What exists in the material ontology is just a separable field on spacetime constrained by a simple law. 
  
  What about the GRW Wentaculus? This is a more delicate issue, because of interpretational questions about the nature of stochastic laws. However, if we think of GRW as giving us guidance about which histories are typical, then the initial density matrix together with stochastic dynamics will fix a class of the histories that are typical according to the laws, and any local probabilities can be obtained by conditionalizing the universal history on available records.  

The nomological interpretation is much more attractive on the Wentaculus than on the Mentaculus, because the universal density matrix is guaranteed to be as simple as the PH. The nomological interpretation of the initial density matrix is compatible with both Humeanism and non-Humeanism. In particular, it demonstrates that quantum entanglement need not be a threat to Humean supervenience \citep{chen2018HU}.

\subsection{Statistical Mechanical Probabilities}

Another advantage of the Wentaculus is that there is only one kind of probability left---quantum mechanical probability. The Statistical Postulate (understood either as a probability or typicality measure)  in the Mentaculus becomes redundant, because the Wentaculus allows only one nomologically possible initial quantum state. 

For example, on the Bohmian Wentaculus, the only probability law we need is the distribution postulate of the initial particle configuration. On the Everettian Wentaculus, the only probability corresponds to the weights of the actual branches, which are interpreted decision-theoretically or using self-locating uncertainty. On the GRW Wentaculus, the only probability corresponds to the collapse chances. Hence, the probability map of the universe is entirely based on quantum probabilities \citep{chen2018valia}. This way of reducing the sources of probability is more conservative than the proposals of \cite{albert1994foundations} and \cite{wallace2011logic}.   For the Everettians, this has an additional bonus, to which I turn now. 

\subsection{Strong Determinism}

The elimination of statistical mechanical probability and maximal constraint on the initial density matrix leads to an interesting consequence for Everettians. The Everettian Wentaculus becomes strongly deterministic, in the sense that there is only one nomologically possible history given the fundamental laws. If we take the fundamental laws seriously, then we have eliminated all sources of arbitrariness in the theory, including that of the initial microstate. The theory would have explained everything at the fundamental level, leaving nothing nomologically contingent \citep{chen2022strong}. 

The Everettian Wentaculus, I believe, is the first realistic and simple example of strong determinism. This has implications for thinking about the relevance of strong determinism to our world. Even if one regards Everettian quantum mechanics as the wrong solution to the measurement problem, it would be dogmatic to regard it as impossible, because it may be empirically equivalent to the other quantum theories. Hence, strong determinism may be closer to the actual world than we have imagined. 


For non-Everettian versions of the Wentaculus where the quantum state obeys unitary dynamics, such as the Bohmian Wentaculus, we have strong determinism with respect to the quantum state history. Given the fundamental laws, the history of the quantum state could not have been different. The only nomological contingency comes from that of the initial particle configuration.

\subsection{Nomic Vagueness}

Standard versions of the PH, such as those found in all three versions of the Mentaculus, are best understood as fundamental yet vague laws. Removing their vagueness by picking an exact set of microstates leads to an objectionable kind of arbitrariness, that I call untraceability \citep{chen2018NV}.  The reason is that the PH is a macrostate law that does not directly enter into the micro-dynamical laws. 

We can regard the Initial Projection Hypothesis as an exact law without committing to such arbitrariness in nature. The initial density matrix simultaneously plays the role of the macrostate and the role of the microstate. It spans the entire macrospace---$\mathscr{H}_{PH}$ and yet it also appears in the micro-dynamical laws. Its exact values will make a difference to what there is in spacetime or how it evolves. For example, in versions where the initial density matrix is ontic, it is automatically traceable. In versions where it is nomic, it is still traceable just like a constant of nature. Any slight change in its values will in general affect how things evolve, for example, by making a difference in the Bohmian velocity field or in the mass-density field. 

This has implications for the mathematical expressibility of fundamental laws. The untraceable arbitrariness is a cost for eliminating fundamental nomic vagueness by fiat, but the Wentaculus takes that cost away. Hence, the Wentaculus enables us to maintain the exactness of fundamental laws without any cost. 

\subsection{Theoretical Unity}

With Density-Matrix Mentaculus, the Wentaculus offers more theoretical unity than Wave-Function Mentaculus. In Everettian and GRW cases, there is more kinematical unity. Even if the universe is in a pure state, most quasi-isolated subsystems do not have pure states.

The Wentaculus (as well as Density-Matrix Mentaculus) provides additional dynamical unity in the Bohmian case.  Suppose the universe is partitioned into a system $S_1$ and its environment $S_2$.   Since there are no actual spins to plug into the spin indices of the wave function, we cannot always define conditional wave functions in an analogous way. Still, we can follow \cite{durr2005role} to define a \emph{conditional density matrix} for $S_1$, by plugging in the actual configuration of $S_2$ and tracing over the spin components in the wave function associated with $S_2$. The conditional density matrix for $S_1$ is defined as:
\begin{equation}\label{WCond}
{W_{cond}}_{s_1'}^{s_1}(q_1, q_1') = \frac{1}{N} \sum_{s_2} \Psi^{s_1 s_2} (q_1, Q_2) \Psi_{s_1 s_2}^* (q_1', Q_2),
\end{equation}
with the normalizing factor:
$N = \int_{\mathcal{Q}_1} dq_1 \sum_{s_1 s_2} \Psi^{s_1 s_2} (q_1, Q_2) \Psi_{s_1 s_2}^* (q_1', Q_2).$
Even if the universe is in a pure state, the configurations of most subsystems are  guided by mixed states according to W-BM.


\section{Objections and Replies}

\subsection{Reliability of Records}

\cite{AlbertLPT} raises two worries about the Wentaculus. Here I address his  worry about the reliability of records on the Bohmian Wentaculus. In \S5.2, I address his worry about time-translation invariance.   

\textbf{Objection:} Suppose we have two momentum eigenstates, one with momentum $+1$ (moving uniformly to the right) and the other $-1$ (moving uniformly to the left). On the Bohmian Mentaculus, one of them is the actual quantum state guiding the particle. The particle will be either moving to the left or to the right. However, on the Bohmian Wentaculus, if we regard the equal mixture of the two  as the fundamental density matrix, the particle guided by this density matrix will be ``entirely, and permanently, and with certainty, at rest.''\footnote{Albert's thought experiment relies on certain idealizations about momentum eigenstates. We usually require the particle configuration to be guided by square-integrable wave functions or density matrices with finite traces, which do not include momentum eigenstates. However, as Sheldon Goldstein points out to me, the example can be fixed by considering momentum eigenstates defined on a closed circle instead of on an infinite line.}   
Nevertheless, when we measure the particle, the record may indicate that the particle is moving.  Hence,  that is a troubling \textit{mismatch between the experimental outcome and the state of the particle} on the Bohmian Wentaculus: the particle is not moving but the record says it is. The mismatch undermines our confidence in the reliability of records in a way analogous to the worry about Bell's Everett (?) theory (\citeyear{bell1981}).

\textbf{Reply:} In a single-particle universe, there are no macroscopic records or human experiences. In a many-particle universe, things become more realistic and interestingly different. To analyze measurements and records, we need to consider the full physical setup of the subsystem and the environment. 

Does the particle move when we include the environment? The answer is yes. Before measurement, the conditional density matrix of the particle, obtained by plugging  the environmental configuration (where the recording device is in a ready state) into the universal density matrix, is an equal mixture of the ``+1'' momentum eigenstate and the ``-1'' momentum eigenstate. 
After measurement, the particle is measured and suppose the record indicates that it is moving to the left. Plugging the new environmental configuration into the universal density matrix, the conditional density matrix of the particle is approximately the ``-1'' momentum eigenstate, because the two parts of the universal density matrix have become macroscopically disjoint and decohered, with the configuration sitting in the ``-1'' part.  Hence, the particle is uniformly moving to the left. There is no mismatch between the record and the state of the particle. The reply here is similar to Bohm's reply (\citeyear{bohm1953})  to Einstein (1953).\footnote{In response to Einstein's worry about the particle in a box of length $L$ with a real-valued wave function $\psi(x)=A sin \frac{2\pi n x}{L}$, Bohm points out that ordinary Bohmian mechanics does not ``contradict any known \textit{experimental} facts,'' because  when we carry out a ``momentum measurement,'' the  wave function (of a stationary state) is transformed and the particle starts to move, even though its original momentum is exactly zero. 
} We should have no less confidence in the reliability of records on the Bohmian Wentaculus than on the Bohmian Mentaculus.

\subsection{Time-Translation Invariance}

Albert's second worry targets the nomological interpretation of the quantum state on the Wentaculus. 
The  reason for postulating an initial universal quantum state, on the nomological interpretation, is to constrain how material objects move in spacetime.  Hence, the universal quantum state is not some material object with  its own independent dynamics. The only things that move and change should be particles (or fields) in physical spacetime, with (\ref{N4})  being the only fundamental dynamical law on the Bohmian Wentaculus and (\ref{Nmxt}) on the Everettian Wentaculus.   Since the right hand sides of (\ref{N4}) and (\ref{Nmxt}) change their functional forms over time, they are not time-translation invariant. 

\textbf{Objection:}  The Wentaculus on the nomological interpretation of the quantum state is fundamentally non-time-translation-invariant, but the world described by such a theory is phenomenologically time-translation-invariant. The theory is ``divided against itself'' \cite[p.28]{AlbertLPT}.  

\textbf{Reply:} First, it is hard to see why this is a bug from Albert's own perspective. For example,  Bohmian mechanics is not Lorentz invariant, but the phenomenological world described by such a theory is.  Albert's high-dimensional ontological interpretation of the wave function tells us the world is fundamentally $3N$-dimensional (with $N \approx 10^{80}$), while the phenomenological world described by such a theory is $3$-dimensional. Fundamental reality may not be an exact image of our phenomenological world. Sometimes theoretical reasons take us to surprising conclusions about fundamental reality. As long as there is a natural and simple explanation of the phenomena from the fundamental, the theory is not self-undermined. 

Second,  there is a formal and technical sense in which we can recover a set of non-fundamental laws of the motion that are time-translation invariant. One can define a universal quantum state $W_t$ from the fundamental laws (the Initial Projection Hypothesis and the von Neumann equation). With respect to this derivative object $W_t$, the particle motion will be time-translation invariant. This explains why there exists a predictive recipe that is time-translation invariant; the violation at the fundamental level makes no practical difference.   

Third,  I regard invariances and symmetries as only defeasible indicators for simplicity, and the lack thereof as defeasible indicators for complexity.  Overall simplicity is something we should strive for. Does the violation of time-translation invariance render the Wentaculus more complicated than the Mentaculus? No; in fact the theory becomes simpler because of it. The violation of invariance in this case indicates nothing useful, because we already know we have a simpler theory. 

Finally,  the non-invariance may be regarded not as a cost but an advantage of the theory, as manifestation of a deeper unity. In the Mentaculus, the theory is not fundamentally time-translation-invariant, because PH applies only at a particular time. However, we can still understand a sense in which the Mentaculus is  time-translation-invariant. We can separate the dynamics from the boundary condition constraint; the dynamics is invariant even though the boundary condition is not. But in the Wentaculus, on the nomological interpretation, there is no such clean separation. The two are genuinely intertwined and unified into a single law. 


\subsection{Ontological Redundancy}
The objection from ontological redundancy has come up in conversations. (\cite[p.399]{wallace2012emergent} mentions but does not necessarily endorse an objection like this.) 

\textbf{Objection:} The Wentaculus requires us to accept density matrix realism, which leads to ``a major expansion of our ontology, from admitting only pure states, to admitting also mixed states.'' And this seems problematic and  unjustified. 

\textbf{Reply:}  Density Matrix Realism does not have a larger ontology (about what actually exists) than Wave Function Realism. In fact, both frameworks postulate that there is exactly one actual quantum state of the universe. Their difference is a modal one, having to do with which states are possible. However, the possibility here is stronger than metaphysical possibility but potentially weaker than nomological possibility. They impose substantive constraints on what kind of states the universe can have. There can be additional fundamental laws of nature, such as the Initial Projection Hypothesis, that further limit such possibilities.  

Perhaps what is behind this worry is the following parsimony principle:
\begin{description}
  \item[Parsimony of Nomological Possibilities] All else being equal, we should prefer theories with smaller sets of nomological possibilities. 
\end{description}
Ironically, this principle works against the objection, because it supports the Wentaculus over the Mentaculus. 
 The Wentaculus is compatible with exactly one nomologically possible initial quantum state, while the Mentaculus is compatible with infinitely many. Hence, by the parsimony of nomological possibilities,  all else being equal, we should prefer the Wentaculus to the Mentaculus.   From the perspective of the Wentaculus, it is the Mentaculus that leads to a major expansion of our nomological possibilities. 


\subsection{The Classical Analogue}
The final  worry is hard to articulate but important to address, because many philosophers of physics have raised this objection in conversations.

\textbf{Objection:}  The same ``trick'' can be played in the classical context. This means that all the advantages of the Wentaculus are too easy to achieve and therefore trivial. On first glance, the suggested maneuver is to take the ``probability distribution'' ($\rho$)  as ``ontic'' or ``nomic.'' The same thing can presumably be done in the classical context (see \cite{McCoySMS} for an example), where the probability distribution on phase space can be given a similarly ontic or nomic interpretation, thus avoiding the problems in the classical domain as well. If that is possible, it seems to show that either we have proven too much, or that it does not depend on the details of quantum theory. 

\textbf{Reply:}  It is much less natural to give an ontic or nomic interpretation of the  probability distribution in classical statistical mechanics. If we use the same idea in the classical domain, we will get a many-worlds version of classical mechanics or lose determinism. The classical probability distribution $\rho$ plays no dynamical role (unlike the density matrix in the W-quantum theories). Since $\rho$ follows the Hamiltonian dynamics, it will in general be supported on many macroscopically distinct regions on phase space. If we reify $\rho$ as ontic and do not modify the dynamics,  we arrive at a many-worlds theory for classical mechanics. If we modify the dynamics to introduce objective ``collapses'' of $\rho$ that take it to some ``branch'' of $\rho$, it will look much more artificial and complex than the original deterministic classical theory. In contrast, on each of the three interpretations of QM, the artificial effects do not arise on the Wentaculus. The Bohmian version remains deterministic (and single-world), the GRW version remains stochastic (and single-world), while the Everettian / many-worlds version is still deterministic.  On the other hand, even if a classical extension of our maneuver is possible, it is unclear how it makes the quantum case trivial, since presumably both require different choices of the ontology and the dynamics. 

\section{Conclusion}


The Wentaculus is an attractive picture of quantum mechanics in a time-asymmetric universe. It is a coherent theory, with arguably a better balance of theoretical virtues than the standard picture. It illuminates the differences between Density Matrix Realism and Wave Function Realism, and displays the advantages of permitting fundamental mixed states. It has implications for our discussions about  laws, chance, randomness, symmetries, vagueness,  determinism, and the quantum reality. If the Wentaculus is correct,  then solutions to the puzzles of time's arrow and quantum ontology are deeply related.  Nature is so unified that we can solves both problems with one key.





\bibliography{test}


\end{document}